\documentclass[12pt,a4paper]{article}
\usepackage{amsmath,amssymb,
booktabs,bm
}
\usepackage{graphicx}
\usepackage{epstopdf}
\usepackage{amsmath,amssymb}

\allowdisplaybreaks[1]
\numberwithin{equation}{section}

\renewcommand{\thefootnote}{\fnsymbol{footnote}}

\def\Lim{\qopname\relax\@empty{lim}\limits}
\let\lim = \Lim

\def\sup{\mathop{\operator@font sup}\limits}
\def\inf{\mathop{\operator@font inf}\limits}
\def\max{\mathop{\operator@font max}\limits}
\def\min{\mathop{\operator@font min}\limits}
\def\prod{\mathop{\mathchoice{\textstyle\prod@}{\textstyle\prod@}{%
      \scriptstyle\prod@}{\scriptscriptstyle\prod@}}\limits}
\def\coprod{\mathop{\mathchoice{\textstyle\coprod@}{\textstyle\coprod@}{%
      \scriptstyle\coprod@}{\scriptscriptstyle\coprod@}}\limits}
\def\bigcap{\mathop{\mathchoice{\textstyle\bigcap@}{\textstyle\bigcap@}{%
      \scriptstyle\bigcap@}{\scriptscriptstyle\bigcap@}}\limits}
\def\bigcup{\mathop{\mathchoice{\textstyle\bigcup@}{\textstyle\bigcup@}{%
      \scriptstyle\bigcup@}{\scriptscriptstyle\bigcup@}}\limits}
\def\Int{\displaystyle\intop\ilimits@}
\def\bigoplus{\mathop{\mathchoice{\textstyle\bigoplus@}{%
      \textstyle\bigoplus@}{\scriptstyle\bigoplus@}{%
      \scriptscriptstyle\bigoplus@}}\limits}
\newcommand{\siml}[0]{\hspace{0.3em}\raisebox{0.4ex}
	{$<$}\hspace{-0.7em}\raisebox{-0.7ex}{{\footnotesize $\sim$}}\hspace{0.4em}}

\def\slashchar#1{\setbox0=\hbox{$#1$}	
\dimen0=\wd0				
\setbox1=\hbox{/} \dimen1=\wd1		
\ifdim\dimen0>\dimen1			
\rlap{\hbox to \dimen0{\hfil/\hfil}}	
#1					
\else 					
\rlap{\hbox to \dimen1{\hfil$#1$\hfil}}	
/					
\fi}
\begin{document}
\begin{titlepage}
 
 \begin{flushright}
 \end{flushright}
 
 \vspace{1ex}
 
 \begin{center}
  
  {\Large \bf Influence of interactions on particle production induced by time-varying mass terms}
  
  \vspace{3ex}
  
  {\large Seishi Enomoto\footnote{e-mail: Seishi.Enomoto@fuw.edu.pl},
  Olga Fuksi\'nska\footnote{e-mail: Olga.Fuksinska@fuw.edu.pl}
  and Zygmunt Lalak\footnote{e-mail: Zygmunt.Lalak@fuw.edu.pl}}
  
  \vspace{4ex}
  {\it Institute of Theoretical Physics, University of Warsaw,\\
   ul. Pasteura 5, 02-093 Warsaw, Poland}
  \vspace{6ex}
 
 \end{center}

\begin{abstract}
We have investigated effects of interaction terms on non-perturbative particle production.
It is well known that time-varying masses induce abundant particle production.
In this paper we have shown that it is possible to induce particle production even
if the mass term of a particle species is not varying in time.
Such particles are produced through the interactions with other fields, whose mass terms are varying due to a time-dependent background.
The necessary formalism has been introduced and analytic and numerical calculations have been performed in a simple but illustrative model. 
The rather general result is that the amount of produced particles without time-dependent masses
can be comparable with the particle density produced directly by the varying background
if the strength of interaction terms is reasonably large. 
\end{abstract}

\end{titlepage}

\renewcommand{\thefootnote}{\arabic{footnote}}
\setcounter{footnote}{0}

\section{Introduction}

In a free quantum theory particle numbers of various species are conserved.
However, if there is a time varying background or a source term in the system,
particle number conservation gets broken and particles are produced
by non-perturbative effects.
It is well known that an oscillating electric field causes production of pairs of electrons \cite{Brezin:1970xf},
and time-varying or inhomogeneous metric causes gravitational particle production \cite{Parker:1969au,Ford:1986sy}. 
In this context the period of post-inflationary preheating, when particles coupled to the inflaton field are produced
on the expense of coherently oscillating inflaton background,
is quite interesting for both particle physics and cosmology because the phenomenon not only marks the end of inflation
but also produces matter constituting our observable Universe \cite{Kofman:1994rk,Kofman:1997yn}. In this paper we are interested in non-gravitational particle production with a signficant role of interaction (non-quadratic) terms. 

The basic mechanism of particle production can be explained with the help of a simple model \cite{Kofman:2004yc}:
\begin{equation}
\label{eom}
 \mathcal{L} = \left| \partial \phi \right|^2
  + \frac{1}{2} \left( \partial \chi \right)^2 - \frac{1}{2} g^2 \left| \phi \right|^2 \chi^2,
\end{equation}
where $\phi$ and $\chi$ are respectively a complex and a real bosonic field, and $g$ is a real coupling constant. Lagrangian (\ref{eom}) may for example describe the coupling between inflaton $\phi$ and a Standard Model field $\chi$ resulting in production of visible particles.
In this model $\chi=0$ becomes a special kind of point in the phase space where the other field becomes effectively massless.\footnote{Of course, also $\langle \phi \rangle = 0$ has the same properties.}
Especially, in case of $\langle \chi \rangle =0$, particle vacuum can be realized for an arbitrary value of $\phi$.
Let us consider the case with $\phi$ having a time-dependent vacuum expectation value (vev) $\left< \phi \right>$.
Asymptotically, when quantum effects can be neglected, one can choose a vacuum solution of (\ref{eom}) in the form of uniform motion:
\begin{equation}
 \left< \phi \right> = vt+i\mu,
\end{equation}
where a real $v$ represents a velocity in the $\phi$-space and a real $\mu$ is an impact parameter measuring the distance of the classical trajectory from the (massless) point $\phi=0$.
When this trajectory approaches the massless point within the circle $\left| \left< \phi \right> \right| \siml \sqrt{v/g}$,
which is the non-adiabaticity condition for $\chi$, mass of $\chi$ becomes very small.
As a result, kinetic energy of the background field $\phi$ is transferred to the field $\chi$ and this leads to the production of $\chi$ particles
whose distribution is calculated to be \cite{Kofman:2004yc},\cite{Enomoto:2013mla}
\begin{equation}
 n_{\mathbf{k}} = \frac{\left( g|v|\right)^{3/2}}{(2\pi)^3} e^{-\pi \frac{\mathbf{k}^2 + g^2 \mu^2}{g|v|}} \label{n}
\end{equation}
where $\mathbf{k}$ is a momentum of a $\chi$ particle.
Furthermore, an interesting phenomenon occurs after the initial period of $\chi$ particle production.
Once particles $\chi$ are created, and trajectory of $\left< \phi \right>$ goes away from the massless point,
energy density of $\chi$ particles can be represented in the Lagrangian by a  term which has a meaning of mass multiplied by number density, i.e.
\begin{equation}
 \rho_{\chi} \sim g \left|\left< \phi \right> \right| \int \frac{d^3k}{(2\pi)^3}n_{\mathbf{k}}.
\end{equation}
This implies that a linear potential is established in the field space. 
Due to the force exerted by this term in the Lagrangian, after a while trajectory of $\left< \phi \right>$ may go back towards the origin and $\chi$ particles can be produced again. Finally, after a number of cycles, trajectory $\left< \phi \right>$ may get trapped near the massless point.

Our main goal in this paper is to extend above considerations to the case where interactions between fields are significant and may become a leading effect, for instance in the absence of the effective  mass terms. As an illustrative example we take a simple supersymmetric generalisation of the model (\ref{eom}) with unbroken global supersymmetry. 
In such a case, fermionic particles $\psi_{\phi}$ and $\psi_{\chi}$,
which are superpartners of $\phi$ and $\chi$, are added to the previous theory.
As we will discuss later, if $\phi$ has a vev then $\chi$ and $\psi_{\chi}$ obtain
masses proportional to $\left|\left< \phi \right> \right|$, while $\psi_{\phi}$ stays massless.
It is easy to see that $\chi$ and $\psi_{\chi}$ particles are produced. 
On the other hand, the mass of $\psi_{\phi}$ is not varying with time since it is absent from the Lagrangian.
However, since the theory we consider has more interactions, e.g. $\chi \psi_{\phi} \psi_{\chi}$,
we shall need to understand their effect on particle production.
As a result, we will find that massless particle can in fact be produced due to interactions.

Outline of the paper is as follows. Section 2 presents the general framework used to calculate number density of produced particles in case of existence of the interaction terms and quantum corrections in the theory; general form of Bogoliubov transformation is also described there. Section 3 focuses on application of the method described in Section 2 to the particular supersymmetric model in both analytical and numerical approach. Rather technical Subsection 3.1 defines precisely asymptotic fields and Bogoliubov transformations in our model.

\section{Quantum corrections to particle production} \label{sec:QE}
Before we analyse a specific model, let us formulate the general recipe for calculation of the number of produced particles in the presence of quantum corrections.
At first we define an occupation number of particles with momentum $\mathbf{k}$ as
\begin{equation}
 n_k \equiv \left< 0^{\rm in} \left| a_{\mathbf{k}}^{{\rm out}\dagger}
  a_{\mathbf{k}}^{\rm out} \right| 0^{\rm in} \right>, \label{eq:def_num}
\end{equation}
where $\left| 0^{\rm in} \right>$ is an in-state vacuum, and $a_{\mathbf{k}}^{\rm out}$ is an
annihilation operator defined by an out-state of vaccum.
In order to calculate this quantity, one employs usually the method of Bogoliubov transformation.
With this method one establishes the relation between $a_{\mathbf{k}}^{\rm in}$ and $a_{\mathbf{k}}^{\rm out}$ in the form
\begin{equation}
 a_{\mathbf{k}}^{\rm out} = \alpha_k a_{\mathbf{k}}^{\rm in} + \beta_k a_{\mathbf{k}}^{{\rm in}\dagger},
\end{equation}
where $\alpha_k$ and $\beta_k$ are complex functions of $|\mathbf{k}|$.
However, calculation of the Bogoliubov coefficients becomes tricky when theory has large interaction terms.
Then, because the out-state is affected by other fields via interaction terms,
the $a_{\mathbf{k}}^{\rm out}$ shall contain information about all field operators in the interacting model, not only about its own field operator.

In this section we will  discuss how the Bogoliubov coefficients are calculated in a complete interacting model. 
Let us begin with a real scalar field $\Psi$.
We assume that the scalar field operator satisfies the following commutation relation
\begin{equation}
 \left[ \Psi(t, \mathbf{x}), \dot{\Psi}(t, \mathbf{y}) \right]
  = i\delta^3(\mathbf{x}-\mathbf{y}) \label{eq:Psi_com_rel1}
\end{equation}
and equation of motion is of the form
\begin{equation}
 0 = \left( \partial^2 + M^2(x) \right) \Psi(x) + J(x), \label{eq:Psi}
\end{equation}
where $M$ is a mass of $\Psi$ which is a classical number (i.e. not an operator),
and $J$ is a source term which is an operator consisting of $\Psi$ or other fields.
In general $M$ may be a function of space-time coordinates $x$, but here we assume only time-dependence for simplicity: $M=M(x^0)$.

Equation (\ref{eq:Psi}) has a formal solution known as {\it Yang-Feldman equation} \cite{Yang:1950vi} given as
\begin{equation}
 \Psi(x) = \sqrt{Z} \Psi^{\rm as}(x) - i Z \int_{t^{\rm as}}^{x^0} dy^0 \int d^3y
  \left[ \Psi^{\rm as}(x), \Psi^{\rm as}(y) \right] J(y), \label{eq:Psi_YF}
\end{equation}
where $\Psi^{\rm as}$ is called an asymptotic field which satisfies free field equation
\begin{equation}
 0 = \left( \partial^2 + M^2 \right) \Psi^{\rm as} \label{eq:Psi_as_EOM} 
\end{equation}
and $Z$ denotes the field renormalization constant.\\
In general, source term $J$ is a function of interacting field operator.
However, one can obtain a sequentially-iterated result from this formal solution substituting (\ref{eq:Psi_YF}) back to $J$ again. In particular, if we take $x^0=t^{\rm as}$ , (\ref{eq:Psi_YF}) becomes
\begin{equation}
 \Psi(t^{\rm as}, \mathbf{x}) = \sqrt{Z} \Psi^{\rm as}(t^{\rm as},\mathbf{x}).
\end{equation} 
Therefore $t^{\rm as}$ denotes a specific point of time when an interacting field $\Psi$ can be regarded as a free field $\Psi^{\rm as}$.
Using this relation one can represent the out-field (defined at $t^{\rm as} = +\infty \equiv t^{\rm out}$)
by the in-field (defined at $t^{\rm as} = -\infty \equiv t^{\rm in}$) as follows:
\begin{equation}
 \Psi^{\rm out}(t^{\rm out},\mathbf{x}) = \Psi^{\rm in}(t^{\rm out}, \mathbf{x})
  - i \sqrt{Z} \int d^4y
  \left[ \Psi^{\rm in}(t^{\rm out}, \mathbf{x}), \Psi^{\rm in}(y) \right] J(y). \label{eq:Psi_in-out-rel}
\end{equation}
One can see that out-field operator depends not only on the in-field but also on the source term which contains interaction effects.

Since (\ref{eq:Psi_as_EOM}) is a free field equation of motion,
solution of $\Psi^{\rm as}$ can be expanded into wave functions as
\begin{equation}
 \Psi^{\rm as}(x) = \int \frac{d^3k}{(2\pi)^3} e^{i \mathbf{k \cdot x}}
  \left[ \Psi_k^{\rm as}(x^0) a_{\mathbf{k}}^{\rm as}
  + \Psi_k^{{\rm as}*}(x^0) a_{\mathbf{-k}}^{{\rm as}\dagger} \right], \label{eq:Psi_as_expand}
\end{equation}
where $a_{\mathbf{k}}^{\rm as}$ 
is an annihilation operator which, following equation (\ref{eq:Psi_as_EOM}), satisfies commutation relation
\begin{equation}
 \left[ a_{\mathbf{k}}^{\rm as}, a_{\mathbf{k'}}^{{\rm as}\dagger} \right]
  = (2\pi)^3 \delta^3(\mathbf{k-k'}), \qquad ({\rm other \: relations}) = 0 \label{eq:Psi_com_rel2}
\end{equation}
and $\Psi_k^{\rm as}$ (where $k \equiv |\mathbf{k}|$)
is a time dependent wave function which satisfies
\begin{equation}
 0 = \ddot{\Psi}_k^{\rm as} + (\mathbf{k}^2 + M^2) \Psi_k^{\rm as}. \label{eq:wave_func_EOM}
\end{equation}
Using commutation relations (\ref{eq:Psi_com_rel1}) and (\ref{eq:Psi_com_rel2}),
one can obtain a time-independent inner product relation \footnote{Time independence can be shown if one takes time derivative of equation (\ref{eq:inner_pro}) and then uses equation of motion.} for the wave function:
\begin{equation}
 \dot{\Psi}_k^{{\rm as}*} \Psi_k^{\rm as} - \Psi_k^{{\rm as}*} \dot{\Psi}_k^{\rm as} = i/Z. \label{eq:inner_pro}
\end{equation} 
With this formula one can extract $a_{\mathbf{k}}^{\rm as}$ from $\Psi^{\rm as}$ as follows:
\begin{equation}
 a_{\mathbf{k}}^{\rm as} = -iZ \int d^3x e^{-i\mathbf{k \cdot x}}
  \left( \dot{\Psi}_k^{{\rm as}*} \Psi^{\rm as} - \Psi_k^{{\rm as}*} \dot{\Psi}^{\rm as} \right).
  \label{ann}
\end{equation}
It means that a component along the $\Psi_k^{\rm as}$-direction
 is projected out of $\Psi^{\rm as}$ by means of the inner product (\ref{eq:inner_pro}).
Relation (\ref{ann}) for out-states can be obtained using (\ref{eq:Psi_in-out-rel}) as
\begin{eqnarray}  \label{eq:BBT}
a_{\mathbf{k}}^{\rm out} &=& \alpha_k a_{\mathbf{k}}^{\rm in}
  + \beta_k a_{-\mathbf{k}}^{{\rm in}\dagger}  \\
 & & - Z^{3/2} \int d^3x e^{-i\mathbf{k \cdot x}} \int d^4y 
  \left[ \dot{\Psi}_k^{{\rm out}*} (x) \Psi^{\rm in} (x) - \Psi_k^{{\rm out}*} (x) \dot{\Psi}^{\rm in} (x), \Psi^{\rm in} (y) \right] J(y) \nonumber  
    \end{eqnarray}
or
  \begin{eqnarray}
a_{\mathbf{k}}^{\rm out} &=& \alpha_k a_{\mathbf{k}}^{\rm in}
  + \beta_k a_{-\mathbf{k}}^{{\rm in}\dagger} \nonumber \\
 & & -i\sqrt{Z} \int d^4x e^{-i\mathbf{k \cdot x}}
  \left( - \beta_k \Psi_k^{\rm in}(x^0) + \alpha_k \Psi_k^{{\rm in}*}(x^0) \right) J(x), \label{eq:BT_a}  
  \end{eqnarray}
where time-independent coefficients $\alpha_k$ and $\beta_k$ are defined as
\begin{eqnarray}
 \alpha_k & \equiv & -iZ \left( \dot{\Psi}_k^{{\rm out}*} \Psi_k^{\rm in}
  - \Psi_k^{{\rm out}*} \dot{\Psi}_k^{\rm in} \right), \\
 \beta_k & \equiv & -iZ \left( \dot{\Psi}_k^{{\rm out}*} \Psi_k^{{\rm in}*}
  - \Psi_k^{{\rm out}*} \dot{\Psi}_k^{{\rm in}*} \right).
\end{eqnarray}
From these relations one learns how much of $\Psi_k^{\rm out}$ or $\Psi_k^{{\rm out}*}$ can be found in $\Psi_k^{\rm in}$.
Thus one can regard $\alpha_k$ and $\beta_k$ as the Bogoliubov coefficients, which
satisfy the following properties:
\begin{eqnarray}
 \Psi_k^{\rm out} &=& \alpha_k^* \Psi_k^{\rm in} - \beta_k^* \Psi_k^{{\rm in}*}, \\
 \Psi_k^{\rm in} &=& \alpha_k \Psi_k^{\rm out} + \beta_k^* \Psi_k^{{\rm out}*},
\end{eqnarray}
and proper normalization
\begin{equation}
 |\alpha_k|^2 - |\beta_k|^2 = 1.
\end{equation}
Hence, formulae (\ref{eq:BBT}) and (\ref{eq:BT_a}) represent a Bogoliubov transformation law generalized to the case of interacting theory. First two terms in these expressions correspond to the usual Bogoliubov transfomation law arising from the time (or space) variation of masses in the Lagrangian and has nothing to do with the source operator $J(x)$. The third term describes the effects coming from interaction terms (quantum corrections).

Finally, expression (\ref{eq:BT_a}) enables us to establish the formula for the occupation number of produced particles in case of interacting theory
\begin{eqnarray} \label{eq:num_Psi}
 n_k & \equiv & \left< 0^{\rm in} \left| a_{\mathbf{k}}^{{\rm out}\dagger}
  a_{\mathbf{k}}^{\rm out} \right| 0^{\rm in} \right> \\ \nonumber
 &=& \left| \left( \beta_k a_{-\mathbf{k}}^{{\rm in}\dagger}
  -i\sqrt{Z} \int d^4x e^{-i\mathbf{k \cdot x}}
  \left( - \beta_k \Psi_k^{\rm in} + \alpha_k \Psi_k^{{\rm in}*} \right)
   J \right) \left| 0^{\rm in} \right> \right|^2 \\ \nonumber 
 &=& \left\{
 \begin{array}{ccll}
  V \cdot |\beta_k|^2 & + & \cdots & (\beta_k \neq 0) \\ \nonumber
  0 & + & Z \left| \int d^4x e^{-i\mathbf{k \cdot x}}
  \Psi_k^{{\rm in}*} J \left| 0^{\rm in} \right> \right|^2 & (\beta_k = 0) \nonumber 
 \end{array}
 \right. ,
\end{eqnarray}
where $V$ is the volume of the system.
This result shows that because of the interaction effects particle production can be induced in both cases, $\beta_k=0$ and $\beta_k \neq 0$, even though the leading term of $n_k$ for $\beta_k=0$ is equal to zero.\\
Analogous results hold for fermions, with a different commutation relation.

\section{Application to a supersymmetric toy model}
In this section we consider production of particles in a specific model. For the sake of simplicity we assume unbroken supersymmetry and the superpotential
\begin{equation}
 W = \frac{1}{2} g \Phi X^2 ,
\end{equation}
where $\Phi$ and $X$ are chiral superfields which interact with a coupling strength $g$.
Corresponding interaction lagrangian reads
\begin{equation}
 \mathcal{L}_{\rm int} = -g^2 |\phi|^2 |\chi|^2 - \frac{1}{4} g^2 |\chi|^4
  - \left( \frac{1}{2} g \phi \psi_{\chi} \psi_{\chi} + g \chi \psi_{\phi} \psi_{\chi} + h.c. \right), \label{eq:lagrangian}
\end{equation}
where $\phi/\psi_{\phi}$ and $\chi/\psi_{\chi}$ are the scalar/fermion components of $\Phi$ and $X$ supermultiplets respectively.
The massless point is  $\chi=\psi_{\psi}=\psi_{\chi}=0$ and $\phi$ can have an arbitrary value there.
Therefore if $\phi$ has a vev $\left< \phi \right> \equiv \left< 0^{\rm in} \left| \phi \right| 0^{\rm in} \right>$,
then $\chi$ and $\psi_{\chi}$ are given a mass $g\left<\phi\right>$ but as we quantize a deviation from the vev, $\tilde{\phi} \equiv \phi - \left< \phi \right>$, $\tilde{\phi}$ and $\psi_{\phi}$ remain massless.

There are two simplifications when one restricts oneself to this toy model. The first one is that we can calculate number of produced particles without worrying about mass counterterms due to supersymmetry.
The second is that this is a minimal supersymmetric model which contains
not only fields with time-varying masses but also massless fields. These facts indicate that the Bogoliubov coefficient $\beta_k$ for massive fields ($\chi, \psi_{\chi}$) becomes non-zero,
on the other hand, $\beta_k=0$ for the massless fields ($\tilde{\phi}, \psi_{\phi}$).
As shown in the previous section, it is possible to produce these massless particles even if $\beta_k=0$ due to interaction effects.
Therefore we can compare the produced particle number in all cases between $\beta_k \neq 0$ and $\beta_k=0$.

\subsection{Bogoliubov transformation}
In this, rather technical, section we demonstrate derivation of the Bogoliubov transformation laws for all component fields present in the Lagrangian (\ref{eq:lagrangian}).
The procedure follows steps outlined in Section \ref{sec:QE}.

From (\ref{eq:lagrangian}) one obtains a set of usual equations of motion:
\begin{eqnarray}
 0 &=& \partial^2 \phi + J_{\phi}^{\dagger}, \label{eq:phi} \\
 0 &=& \left( \partial^2 + g^2 \left| \left< \phi \right> \right|^2 \right) \chi + J_{\chi}^{\dagger}, \label{eq:chi} \nonumber \\
 0 &=& i\bar{\sigma}^{\mu} \partial_{\mu} \psi_{\phi} - J_{\psi_{\phi}}^{\dagger}, \label{eq:psi_phi} \nonumber \\
 0 &=& i\bar{\sigma}^{\mu} \partial_{\mu} \psi_{\chi}
   - g \left< \phi^{\dagger} \right> \psi_{\chi}^{\dagger} - J_{\psi_{\chi}}^{\dagger}, \label{eq:psi_chi}\nonumber
\end{eqnarray}
with source terms given as
\begin{eqnarray}
 J_{\phi}^{\dagger} & \equiv & g^2 |\chi|^2 \phi + \frac{1}{2} g \psi_{\chi}^{\dagger} \psi_{\chi}^{\dagger}, \label{source} \\
 J_{\chi}^{\dagger} & \equiv & g^2 \left(|\phi|^2 - \left| \left< \phi \right> \right|^2 \right) \chi
  + \frac{1}{2} g^2 |\chi|^2 \chi + g \psi_{\chi}^{\dagger} \psi_{\phi}^{\dagger}, \nonumber \\
 J_{\psi_{\phi}}^{\dagger} & \equiv & g \chi^{\dagger} \psi_{\chi}^{\dagger}, \nonumber \\
 J_{\psi_{\chi}}^{\dagger} & \equiv & g \left( \phi^{\dagger} - \left< \phi^{\dagger} \right> \right) \psi_{\chi}^{\dagger}
  + g \chi^{\dagger} \psi_{\phi}^{\dagger}. \nonumber
\end{eqnarray}
In the above formulae we took into account the possible influence of the vev $\left< \phi \right>$ on $\chi$ and $\psi_{\chi}$ masses.

Following (\ref{source}) we can obtain the Yang-Feldman equations for scalar fields as
\begin{eqnarray}
\phi(x) &=& \sqrt{Z_{\phi}} \phi^{\rm as}(x) - i Z_{\phi} \int_{t^{\rm as}}^{x^0} dy^0 \int d^3y
  \left[ \phi^{\rm as}(x), \phi^{{\rm as}*}(y) \right] J_{\phi}^{\dagger}(y), \label{eq:formal_phi} \\
 \chi(x) &=& \sqrt{Z_{\chi}} \chi^{\rm as}(x) - i Z_{\chi} \int_{t^{\rm as}}^{x^0} dy^0 \int d^3y
  \left[ \chi^{\rm as}(x), \chi^{{\rm as}*}(y) \right] J_{\chi}^{\dagger}(y).
\end{eqnarray}
Analgous equations for fermions read
\begin{eqnarray}
   \psi_{\phi \alpha}(x) &=& \sqrt{Z_{\psi_\phi}} \psi_{\phi \alpha}^{\rm as}(x)
  - i Z_{\psi_\phi} \int_{t^{\rm as}}^{x^0} dy^0 \int d^3y
   \left\{ \psi_{\phi \alpha}^{\rm as}(x), \psi_{\phi \dot{\beta}}^{{\rm as} \dagger}(y) \right\}
   J_{\psi_{\phi}}^{\dagger \dot{\beta}}(y), \\
 \psi_{\chi \alpha}(x) &=& \sqrt{Z_{\psi_\chi}} \psi_{\chi \alpha}^{\rm as}(x)
  - i Z_{\psi_{\chi}} \int_{t^{\rm as}}^{x^0} dy^0 \int d^3y \nonumber \\
  & & \quad \times \left(
   \left\{ \psi_{\chi \alpha}^{\rm as}(x), \psi_{\chi \dot{\beta}}^{{\rm as} \dagger}(y) \right\}
    J_{\psi_{\chi}}^{\dagger \dot{\beta}}(y)
   + \left\{ \psi_{\chi \alpha}^{\rm as}(x), \psi_{\chi}^{{\rm as} \beta}(y) \right\}
    J_{\psi_{\chi} \beta}(y) \right). \label{eq:formal_psi_chi} \nonumber
\end{eqnarray}

Asymptotic fields, by definition, can be regarded as free fields what enables us to expand them into plane waves. Such decomposition for bosons differs only by a vev ($\langle \chi \rangle = 0$ while $\langle \phi \rangle \neq 0$)
\begin{eqnarray}
 \phi^{\rm as} & = & \left< 0^{\rm as} \left| \phi^{\rm as} \right| 0^{\rm as} \right>
  + \int \frac{d^3k}{(2\pi)^3} e^{i\mathbf{k \cdot x}}
   \left( \phi_k^{\rm as} a_{\phi \mathbf{k}}^{+,{\rm as}}
    + \phi_k^{\rm as*} a_{\phi -\mathbf{k}}^{-,{\rm as} \dagger} \right), \label{eq:phi_expand}\\
 \chi^{\rm as} & = & \int \frac{d^3k}{(2\pi)^3} e^{i\mathbf{k \cdot x}}
   \left( \chi_k^{\rm as} a_{\chi \mathbf{k}}^{+,{\rm as}}
    + \chi_k^{\rm as*} a_{\chi -\mathbf{k}}^{-,{\rm as} \dagger} \right), \label{eq:phi_expand}
   \end{eqnarray}
   whereas for fermions we are obliged to take into account also helicity (see Appendix)
    \begin{eqnarray}
 \psi_{\phi}^{\rm as} & = & \int \frac{d^3k}{(2\pi)^3} e^{i \mathbf{k \cdot x}}
  \left( e_{\mathbf{k}}^+ \psi_{\phi k}^{\rm as} a_{\psi_{\phi} \mathbf{k}}^{+,{\rm as}}
   + e_{\mathbf{k}}^- \psi_{\phi k}^{{\rm as}*} a_{\psi_{\phi} \mathbf{-k}}^{-,{\rm as} \dagger} \right),\\
 \psi_{\chi}^{\rm as} & = & \int \frac{d^3k}{(2\pi)^3} e^{i \mathbf{k \cdot x}} \sum_{s=\pm} e_{\mathbf{k}}^s
  \left( \psi_{\chi k}^{(+)s,{\rm as}} a_{\psi_{\chi} \mathbf{k}}^{s,{\rm as}}
   - s e^{-i\theta_{\mathbf{k}}}\psi_{\chi k}^{(-)s,{\rm as} *}
   a_{\psi_{\chi} \mathbf{-k}}^{s,{\rm as}\dagger} \right), \label{eq:psi_chi_expand}
\end{eqnarray}
Here superscript "as" denotes the wave functions satisfying the following set of equations of motion
\begin{eqnarray}
 0 &=& \ddot{\phi}_k^{\rm as} + \mathbf{k}^2 \phi_k^{\rm as}, \label{eq:phi_k_as} \\
 0 &=& \ddot{\chi}_k^{\rm as} + \left( \mathbf{k}^2
  + g^2 \left| \left< \phi \right> \right|^2 \right) \chi_k^{\rm as},  \label{eq:chi_k_as}\nonumber\\
 0 &=& \dot{\psi}_{\phi k}^{\rm as} + i|\mathbf{k}| \psi_{\phi k}^{\rm as}, \label{eq:psi_phi_k_as} \nonumber \\
 0 &=& \dot{\psi}_{\chi k}^{(+)s,{\rm as}} + is|\mathbf{k}| \psi_{\chi k}^{(+)s,{\rm as}}
  + ig \left< \phi^{\dagger} \right> \psi_{\chi k}^{(-)h,{\rm as}},  \label{eq:psi_chi_k_p_as} \nonumber\\
 0 &=& \dot{\psi}_{\chi k}^{(-)s,{\rm as}} - is|\mathbf{k}| \psi_{\chi k}^{(-)s,{\rm as}}
  + ig \left< \phi \right> \psi_{\chi k}^{(+)h,{\rm as}} \label{eq:psi_chi_k_m_as} \nonumber
\end{eqnarray}
and one has commutation relation for scalars and anticommutation relation for fermions of the form
\begin{equation}
 \left[ a_{\phi \mathbf{k}}^{s,{\rm as}}, a_{\phi \mathbf{k'}}^{r,{\rm as} \dagger} \right] =
 \left[ a_{\chi \mathbf{k}}^{s,{\rm as}}, a_{\chi \mathbf{k'}}^{r,{\rm as} \dagger} \right] =
 (2\pi)^3 \delta^3(\mathbf{k-k'}) \delta^{sr},
\end{equation}
\begin{equation}
 \left\{ a_{\psi_{\phi} \mathbf{k}}^{s,{\rm as}}, a_{\psi_{\phi}  \mathbf{k'}}^{r,{\rm as}\dagger} \right\} =
 \left\{ a_{\psi_{\chi} \mathbf{k}}^{s,{\rm as}}, a_{\psi_{\chi}  \mathbf{k'}}^{r,{\rm as}\dagger} \right\} =
 (2\pi)^3 \delta^3(\mathbf{k-k'}) \delta^{sr}.
\end{equation}
In the above formulae $\pm$ labels distinguish between a particle and antiparticle for scalars and indicate the helicity of fermions.

Using the established commutation relation one can obtain inner product relations obeyed by the components of wave functions for all species:
\begin{eqnarray}
 i &=& Z_{\phi} \left( \dot{\phi}_k^{{\rm as}*} \phi_k^{\rm as}
  - \phi_k^{{\rm as}*} \dot{\phi}_k^{\rm as} \right), \\
 i &=& Z_{\chi} \left( \dot{\chi}_k^{{\rm as}*} \chi_k^{\rm as}
  - \chi_k^{{\rm as}*} \dot{\chi}_k^{\rm as} \right), \nonumber \\
  \end{eqnarray}
  for scalars and
  \begin{eqnarray}
   1 &=& Z_{\psi_{\phi}} \left| \psi_{\phi k}^{\rm as} \right|^2, \\
 1 &=& Z_{\psi_{\chi}} \left( \left| \psi_{\chi k}^{(+) s,{\rm as}} \right|^2
  + \left| \psi_{\chi k}^{(-) s,{\rm as}} \right|^2 \right), \nonumber
\end{eqnarray}
for fermions.

With these products one can extract the expressions for annihilation operators using respective Yang-Feldman equations with $t^{\rm as}=t^{\rm out}\equiv+\infty$:
\begin{eqnarray}
 a_{\phi \mathbf{k}}^{+,{\rm out}} &=& -iZ \int d^3x e^{-i\mathbf{k \cdot x}}
  \left[ \dot{\phi}_k^{{\rm out}*} \left( \phi^{\rm out}
   - \left< 0^{\rm out} \left| \phi^{\rm out} \right| 0^{\rm out} \right> \right) \right. \nonumber \\
  & & \qquad \qquad \qquad \qquad \left. - \phi_k^{{\rm out}*} \left( \dot{\phi}^{\rm out}
   - \left< 0^{\rm out} \left| \dot{\phi}^{\rm out} \right| 0^{\rm out} \right> \right) \right], \label{eq:a_phi} \\
 a_{\phi \mathbf{-k}}^{-,{\rm out}\dagger} &=& +iZ \int d^3x e^{-i\mathbf{k \cdot x}}
  \left[ \dot{\phi}_k^{\rm out} \left( \phi^{\rm out}
   - \left< 0^{\rm out} \left| \phi^{\rm out} \right| 0^{\rm out} \right> \right) \right. \nonumber \\
  & & \qquad \qquad \qquad \qquad \left. - \phi_k^{\rm out} \left( \dot{\phi}^{\rm out}
   - \left< 0^{\rm out} \left| \dot{\phi}^{\rm out} \right| 0^{\rm out} \right> \right) \right], \nonumber
\\
 a_{\chi \mathbf{k}}^{+,{\rm out}} &=& -iZ_{\chi} \int d^3x e^{-i\mathbf{k \cdot x}}
  \left( \dot{\chi}_k^{{\rm out}*} \chi^{\rm out} - \chi_k^{{\rm out}*} \dot{\chi}^{\rm out} \right), \nonumber \\
 a_{\chi \mathbf{-k}}^{-,{\rm out}\dagger} &=& +iZ_{\chi} \int d^3x e^{-i\mathbf{k \cdot x}}
  \left( \dot{\chi}_k^{\rm out} \chi^{\rm out} - \chi_k^{\rm out} \dot{\chi}^{\rm out} \right), \nonumber
\\
 a_{\psi_{\phi} \mathbf{k}}^{+,{\rm out}} &=& Z_{\psi_{\phi}} \int d^3x e^{-i\mathbf{k \cdot x}}
  \psi_{\phi k}^{{\rm out}*} \cdot e_{\mathbf{k}}^{+\dagger} \bar{\sigma}^0 \psi_{\phi}^{\rm out}, \nonumber \\
 a_{\psi_{\phi} \mathbf{-k}}^{-,{\rm out}\dagger} &=& Z_{\psi_{\phi}} \int d^3x e^{-i\mathbf{k \cdot x}}
  \psi_{\phi k}^{\rm out} \cdot e_{\mathbf{k}}^{-\dagger} \bar{\sigma}^0 \psi_{\phi}^{\rm out}, \nonumber
\\
 a_{\psi_{\chi} \mathbf{k}}^{s,{\rm out}} &=& Z_{\psi_{\phi}} \int d^3x e^{-i\mathbf{k \cdot x}}
  \left( \psi_{\chi k}^{(+)s,{\rm out}*} \cdot e_{\mathbf{k}}^{s\dagger} \bar{\sigma}^0 \psi_{\chi}^{\rm out} \right. \nonumber \\
 & & \left. \qquad \qquad \qquad \qquad \qquad + s e^{-i\theta_{\mathbf{k}}} \psi_{\chi k}^{(-)s,{\rm out}}
   \cdot \psi_{\chi}^{{\rm out}\dagger} \bar{\sigma}^0 e_{\mathbf{-k}}^s \right). \label{eq:a_psi_chi} \nonumber
\end{eqnarray}

Also the relation between in and out states, leading to the Bogoliubov transformation, can be found:
\begin{eqnarray}
 \phi^{\rm out}(t^{\rm out}, \mathbf{x}) &=& \phi^{\rm in}(t^{\rm out}, \mathbf{x}) - i \sqrt{Z_{\phi}} \int d^4y
  \left[ \phi^{\rm in}(t^{\rm out}, \mathbf{x}), \phi^{{\rm in}\dagger}(y) \right] J_{\phi}^{\dagger}(y), \\
 \chi^{\rm out}(t^{\rm out}, \mathbf{x}) &=& \chi^{\rm in}(t^{\rm out}, \mathbf{x}) - i \sqrt{Z_{\chi}} \int d^4y
  \left[ \chi^{\rm in}(t^{\rm out}, \mathbf{x}), \chi^{{\rm in}\dagger}(y) \right] J_{\chi}^{\dagger}(y),\nonumber \end{eqnarray}
  for scalars and
  \begin{eqnarray}
   \psi_{\phi \alpha}^{\rm out}(t^{\rm out}, \mathbf{x}) &=& \psi_{\phi \alpha}^{\rm in}(t^{\rm out}, \mathbf{x})
  - i\sqrt{Z_{\psi_\phi}} \int d^4y
   \left\{ \psi_{\phi \alpha}^{\rm in}(t^{\rm out}, \mathbf{x}), \psi_{\phi \dot{\beta}}^{{\rm in} \dagger}(y) \right\}
   J_{\psi_{\phi}}^{\dagger \dot{\beta}}(y), \\
 \psi_{\chi \alpha}^{\rm out}(t^{\rm out}, \mathbf{x}) &=& \psi_{\chi \alpha}^{\rm in}(t^{\rm out}, \mathbf{x})
  - i \sqrt{Z_{\psi_{\chi}}} \int d^4y \left(
   \left\{ \psi_{\chi \alpha}^{\rm in}(t^{\rm out}, \mathbf{x}), \psi_{\chi \dot{\beta}}^{{\rm in} \dagger}(y) \right\}
    J_{\psi_{\chi}}^{\dagger \dot{\beta}}(y) \right. \nonumber \\
  & & \qquad \qquad \qquad \qquad \qquad \left.
   + \left\{ \psi_{\chi \alpha}^{\rm in}(t^{\rm out}, \mathbf{x}), \psi_{\chi}^{{\rm in} \beta}(y) \right\}
    J_{\psi_{\chi} \beta}(y) \right), \nonumber
\end{eqnarray}
for fermions.

Putting together available information we can now derive the following Bogoliubov tranformation rules for scalar $\phi$:
\begin{eqnarray}
 a_{\phi \mathbf{k}}^{+,{\rm out}} &=& \left< a_{\phi \mathbf{k}}^{+,{\rm out}} \right>
   + a_{\phi \mathbf{k}}^{+,{\rm in}} - i\sqrt{Z_{\phi}} \int d^4x e^{-i\mathbf{k \cdot x}} \phi_k^{{\rm out}*}
   \left( J_{\phi}^{\dagger} - \left< J_{\phi}^{\dagger} \right> \right),\\
 a_{\phi \mathbf{-k}}^{-,{\rm out}\dagger} &=& \left< a_{\phi \mathbf{-k}}^{-,{\rm out}\dagger} \right>
   + a_{\phi \mathbf{-k}}^{-,{\rm in}\dagger} + i\sqrt{Z_{\phi}} \int d^4x e^{-i\mathbf{k \cdot x}} \phi_k^{\rm out}
   \left( J_{\phi}^{\dagger} - \left< J_{\phi}^{\dagger} \right> \right), \nonumber
\end{eqnarray}
and for scalar $\chi$:
\begin{eqnarray}
 a_{\chi \mathbf{k}}^{+,{\rm out}}
  &=& \alpha_{\chi k} a_{\chi \mathbf{k}}^{+,{\rm in}}
   + \beta_{\chi k} a_{\chi \mathbf{-k}}^{-,{\rm in} \dagger}
   - i\sqrt{Z_{\chi}} \int d^4x e^{-i\mathbf{k \cdot x}} \chi_k^{{\rm out}*} J_{\chi}^{\dagger},\\
 a_{\phi \mathbf{-k}}^{-,{\rm out}\dagger}
  &=& \beta_{\chi k}^* a_{\chi \mathbf{k}}^{+,{\rm in}}
   + \alpha_{\chi k}^* a_{\chi \mathbf{-k}}^{-,{\rm in} \dagger}
   + i\sqrt{Z_{\chi}} \int d^4x e^{-i\mathbf{k \cdot y}} \chi_k^{\rm out} J_{\chi}^{\dagger}, \nonumber
\end{eqnarray}

In the fermionic case one finds for the $\psi_{\phi}$ field:
\begin{eqnarray}
 a_{\psi_{\phi} \mathbf{k}}^{+,{\rm out}}
  &=& a_{\psi_{\phi} \mathbf{k}}^{+,{\rm in}}
   - i\sqrt{Z_{\psi_{\phi}}} \int d^4x e^{-i\mathbf{k \cdot x}} \psi_{\phi k}^{{\rm out}*}
   \cdot e_{\mathbf{k}}^{+ \dagger} J_{\psi_{\phi}}^{\dagger},\\
 a_{\psi_{\phi} \mathbf{-k}}^{-,{\rm out}\dagger}
  &=& a_{\psi_{\phi} \mathbf{-k}}^{-,{\rm in}\dagger} 
   - i\sqrt{Z_{\psi_{\phi}}} \int d^4x e^{-i\mathbf{k \cdot x}} \psi_{\phi k}^{\rm out}
   \cdot e_{\mathbf{k}}^{- \dagger} J_{\psi_{\phi}}^{\dagger}.\nonumber
\end{eqnarray}
 and for the $\psi_{\chi}$:
\begin{eqnarray}
 a_{\psi_{\chi} \mathbf{k}}^{s,{\rm out}}
  &=& \alpha_{\chi k} a_{\psi_{\chi} \mathbf{k}}^{s,{\rm in}}
   + \beta_{\chi \mathbf{k}} a_{\psi_{\chi} -\mathbf{k}}^{s,{\rm in}\dagger} \nonumber \\
  & & - i\sqrt{Z_{\psi_{\chi}}} \int d^4x e^{-i\mathbf{k \cdot x}} \left( \psi_{\chi k}^{(+)s,{\rm out}*}
   \cdot e_{\mathbf{k}}^{s\dagger} J_{\psi_{\chi}}^{\dagger} \right. \nonumber \\
  & & \qquad \qquad \qquad \qquad \left. + s e^{-i\theta_{\mathbf{k}}} \psi_{\chi k}^{(-)s,{\rm out}*}
   \cdot e_{\mathbf{-k}}^s J_{\psi_{\chi}} \right).
\end{eqnarray}
The Bogoliubov coefficients for massive fields are given by
\begin{eqnarray}
 \alpha_{\chi k} & \equiv & -iZ_{\chi} \left( \dot{\chi}_k^{{\rm out}*} \chi_k^{\rm in}
  - \chi_k^{{\rm out}*} \dot{\chi}_k^{\rm in} \right), \\
 \beta_{\chi k} & \equiv & -iZ_{\chi} \left( \dot{\chi}_k^{{\rm out}*} \chi_k^{{\rm in}*}
  - \chi_k^{{\rm out}*} \dot{\chi}_k^{{\rm in}*} \right), \label{eq:beta_chi} \nonumber
\end{eqnarray}
for scalars and
\begin{eqnarray}
 \alpha_{\psi_{\chi} k}^s & \equiv & Z_{\psi_{\chi}} \left( \psi_{\chi k}^{(+)s,{\rm out}*} \psi_{\chi k}^{(+)s,{\rm in}}
  + \psi_{\chi k}^{(-)s,{\rm out}*} \psi_{\chi k}^{(-)s,{\rm in}} \right), \\
 \beta_{\psi_{\chi} \mathbf{k}}^s & \equiv & -Z_{\psi_{\chi}} se^{-i\theta_{\mathbf{k}}}
  \left( \psi_{\chi k}^{(+)s,{\rm out}*} \psi_{\chi k}^{(-)s,{\rm in}*}
  - \psi_{\chi k}^{(-)s,{\rm out}*} \psi_{\chi k}^{(+)s,{\rm in}*} \right), \nonumber \label{eq:beta_psi_chi}
\end{eqnarray}
for fermions.

It should be noted, that the usual Bogoliubov coefficients are absent in case of massless field $\phi$ and $\psi_{\phi}$.
The reason is that their mass terms do not change with time and their equations of motion can be solved over the whole time interval.  
Therefore the form of wave functions in the in-state and in the out-state is the same,
i.e. $\phi_k^{\rm in}(t)=\phi_k^{\rm out}(t)$ and $\psi_{\phi k}^{\rm in}(t)=\psi_{\phi k}^{\rm out}(t)$, hence the Bogoliubov coefficients $\beta_k$ for $\phi$ and $\psi_{\phi}$ should vanish indeed.

\subsection{Analytical results}
For massive particles $\chi$ and $\psi_{\chi}$, the leading results for occupation numbers are calculated, using the framework described in Section 2, to be:
\begin{eqnarray}
 n_{\chi k} & \equiv & \sum_{s=\pm} \left< 0^{\rm in} \left| a_{\chi \mathbf{k}}^{s,{\rm out}\dagger}
  a_{\chi \mathbf{k}}^{s,{\rm out}} \right| 0^{\rm in} \right>
  \: = \: V \cdot 2 \left| \beta_{\chi k} \right|^2 + \cdots, \label{eq:chi_occu}\\
 n_{\psi_{\chi} k} & \equiv & \sum_{s=\pm} \left< 0^{\rm in} \left| a_{\psi_{\chi} \mathbf{k}}^{s,{\rm out}\dagger}
  a_{\psi_{\chi} \mathbf{k}}^{s,{\rm out}} \right| 0^{\rm in} \right>
  \: = \: V \cdot \sum_{s=\pm} \left| \beta_{\psi_{\chi} \mathbf{k}}^s \right|^2 + \cdots, \label{eq:psi_chi_occu}
\end{eqnarray}
where $V$ is the volume of the system.
The factor 2 in (\ref{eq:chi_occu}) counts degrees of freedom for a complex scalar field. 
If one takes $\left< \phi \right> = vt + i\mu$ as the initial condition,
one obtains
\begin{equation}
 \left| \beta_{\chi \, k} \right|^2 = \left| \beta_{\psi_{\chi} \, \mathbf{k}}^s \right|^2
  \sim e^{-\pi \frac{k^2 + g^2 \mu^2}{g|v|}}, \label{eq:num_chi_psi_chi}
\end{equation}
using the analytical continuation method as in (\ref{n}).
The standard quasi-classical reasoning leads to the same result for Bogoliubov coefficients in case of fermions and in case of bosons. 
As a consequence, we can evaluate the number densities to be
\begin{equation}
 n_{\chi} = n_{\psi_{\chi}} = \int \frac{d^3k}{(2\pi)^3} \frac{n_{\chi k}}{V}
  \: \sim \: 2 \times \frac{(g|v|)^{3/2}}{(2\pi)^3}e^{-\pi g\mu^2/|v|} .
\end{equation}

On the other hand, in the case of production of massless particles $\tilde{\phi}$ and $\psi_{\phi}$ interaction effect in the Bogoliubov transformation formula is the leading contribution, since the respective Bogoliubov coefficient $\beta$ is zero. Up to the lowest power of the coupling $g$, one finds for the massless scalar:
\begin{eqnarray}
 & n_{\phi k} \equiv \sum\limits_{s=\pm} \left< 0^{\rm in} \left|
  \left( a_{\phi \mathbf{k}}^{s,{\rm out}\dagger} - \left< a_{\phi \mathbf{k}}^{s,{\rm out}\dagger} \right> \right)
  \left( a_{\phi \mathbf{k}}^{s,{\rm out}} - \left< a_{\phi \mathbf{k}}^{s,{\rm out}} \right> \right)
  \right| 0^{\rm in} \right> \approx \nonumber \\
 & \approx V \cdot g^2 \int \frac{d^3p}{(2\pi)^3} \left[
  Z_{\phi} Z_{\chi}^2 \left| \int dt \: \phi_k^{\rm out}
  \chi_{|\mathbf{k+p}|}^{\rm in} \chi_p^{\rm in} \cdot g \left< \phi^* \right> \right|^2 \right. + Z_{\phi} Z_{\chi}^2 \left| \int dt \: \phi_k^{\rm out}
  \chi_{|\mathbf{k+p}|}^{\rm in} \chi_p^{\rm in} \cdot g \left< \phi \right> \right|^2 \nonumber \\
 & + \frac{1}{4} Z_{\phi} Z_{\psi_{\chi}}^2 \sum\limits_{s,r,q}
  \left(1+rq\frac{\mathbf{p\cdot}(\mathbf{k+p})}{p|\mathbf{k+p}|}\right) \: \left. \times \left| \int dt \:
  \phi_k^{\rm out} \psi_{\chi |\mathbf{k+p}|}^{(s)r,{\rm in}}
  \psi_{\chi p}^{(s)q,{\rm in}} \right|^2 \right]
  , \label{eq:phi_occu}
  \end{eqnarray}
  and for the massless fermion:
  \begin{eqnarray}
 & n_{\psi_{\phi} k} \equiv \sum\limits_{s=\pm} \left< 0^{\rm in} \left| a_{\psi_{\phi} \mathbf{k}}^{s,{\rm out}\dagger}
  a_{\psi_{\phi} \mathbf{k}}^{s,{\rm out}} \right| 0^{\rm in} \right> \approx \nonumber \\
 & \approx V \cdot g^2 Z_{\chi} Z_{\psi_{\phi}} Z_{\psi_{\chi}} \int \frac{d^3 p}{(2\pi)^3} \sum\limits_{s,r}
  \frac{1}{2} \left(1-sr \frac{\mathbf{k \cdot p}}{kp} \right) \times \left|\int dt \:
  \psi_{\phi k}^{\rm out} \chi_{|\mathbf{k+p}|}^{\rm in} \psi_{\chi p}^{(s)r,{\rm in}} \right|^2 
  . \label{eq:psi_phi_occu}
\end{eqnarray}

\begin{figure}[t]
 \begin{center}
  \includegraphics[scale=0.6]{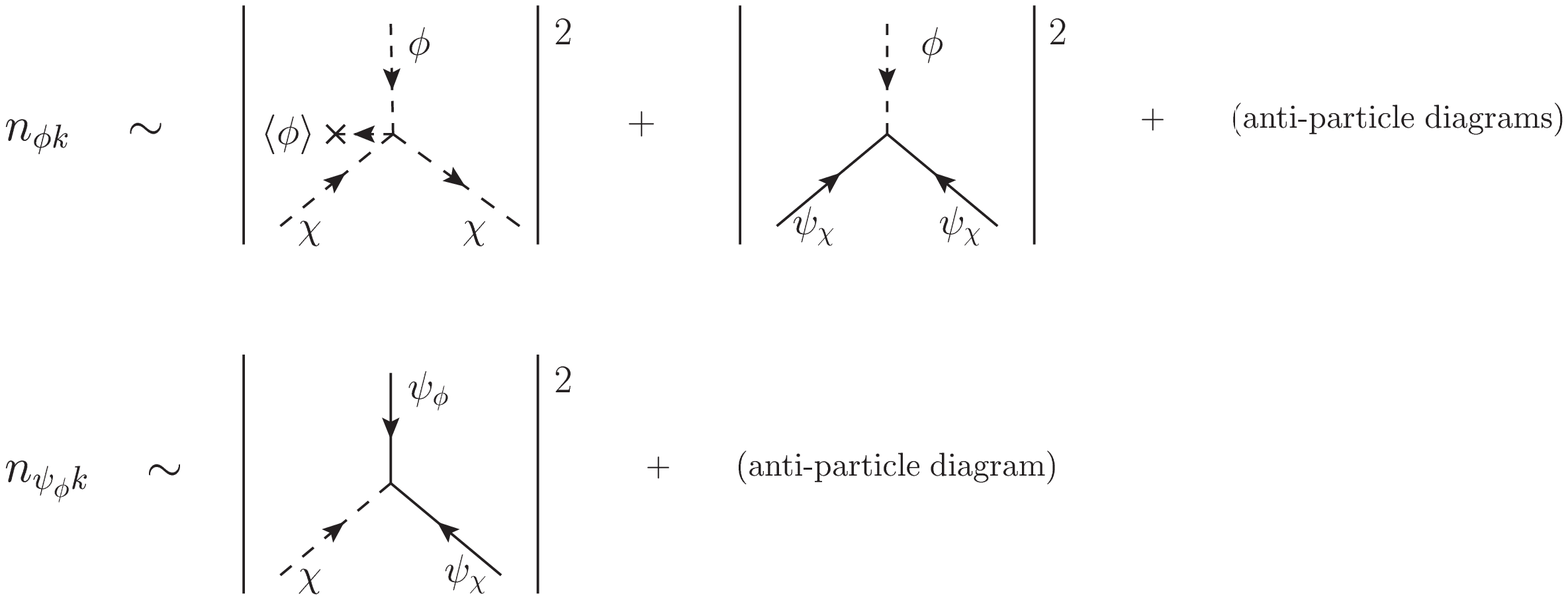}
  \caption{Diagrams corresponding to particle production.
  The momentum integrals and helicity summation are omitted.}
  \label{fig:diagrams}
 \end{center}
\end{figure}

The above formulae look complicated but
we can find the physical meaning  by considering perturbation theory diagrams.
Since the wave function corresponds to an external line of a diagram, the diagrams corresponding to
(\ref{eq:phi_occu}) and (\ref{eq:psi_phi_occu}) are the ones  shown  in the Figure \ref{fig:diagrams}.
Note that these diagrams can be described as "inverse decay" processes.
One might suspect that they are forbidden because of conservation of energy and momentum.
However, these processes are actually possible since the background field is time-dependent
and one needs to take into account energy non-conservation in the diagrams representing quantum processes in this theory.
Therefore, at least from the perturbative point of view, we can conclude that
the main effect in the massless particle production comes from inverse decay processes of massive particles in varying external background.

As far as the loop corrections are concerned, their effect on particle production in case of the weak coupling is
supressed by the growing powers of g multiplied by small numerical loop-factors.
In case of a strong coupling this effect becomes much more subtle
- loop contributions to the propagators and vertices considered here can become large and even growing with time,
as argued in \cite{Akhmedov1} and \cite{Akhmedov2}.
Discussion of potentially large loop effects goes beyond the scope of this paper and shall be addressed elsewhere.

It is difficult to evaluate formulae (\ref{eq:phi_occu}) and (\ref{eq:psi_phi_occu}) analytically
(we will show numerical results in Section \ref{sec:numerical_results}). However,
we can find the behaviour of number densities for massless particles at low momentum (or large vev) if we use rough approximations
\begin{equation}
 \chi_k^{\rm in} \sim \frac{1}{\sqrt{2\omega_k}} e^{- \pi k^2/2g|v|},
 \quad \psi_{\chi k}^{(s)r,{\rm in}} \sim e^{- \pi k^2/2g|v|}, \quad \omega_k \sim g|v||t|.
\end{equation}
This clearly corresponds to the exponential supression of (\ref{eq:num_chi_psi_chi}).
The above approximations simplify  (\ref{eq:phi_occu}) and (\ref{eq:psi_phi_occu}) to the form
\begin{eqnarray}
 n_{\phi k}/V & \sim & g^2 \int \frac{d^3p}{(2\pi)^3} \left| \int dt \frac{1}{\sqrt{2k}} \right|^2 e^{-\pi p^2/g|v|}
  \: \propto \: g^2 \cdot (g|v|)^{3/2} |t|^2 /k, \label{eq:n_phi_k_low_momemntum} \\
 n_{\psi_{\phi} k}/V & \sim & g^2 \int \frac{d^3p}{(2\pi)^3} \left| \int dt \frac{1}{\sqrt{2g|v||t|}} \right|^2 e^{-\pi p^2/g|v|}
  \: \propto \: g^2 \cdot \sqrt{g|v|} |t|. \label{eq:n_psi_phi_k_low_momemntum}
\end{eqnarray}
We will use these results later to make fits to numerical results\footnote{Although the result
(\ref{eq:n_phi_k_low_momemntum}) shows a quadratic time dependence,
which is supported by our numerical results, there are papers which claim a linear time dependence, e.g. see \cite{Akhmedov1,Akhmedov2}.}.

As we mentioned before analytical approach is severely limited in case of particle production in interacting theories. Equations to be solved are those determining $ \langle \phi \rangle$ and in-state wave functions for all available species. For massless ones we can find analytically that
\begin{equation}
 \sqrt{Z_{\phi}} \phi_k^{\rm in}(t) = \frac{1}{\sqrt{2|\mathbf{k}|}} e^{-i|\mathbf{k}|t},
  \qquad \sqrt{Z_{\psi_{\phi}}} \psi_{\phi k}^{\rm in}(t) = e^{-i|\mathbf{k}|t},
\end{equation}
but for the massive case analytical solution is impossible and one is left with WKB type approximate solutions
\begin{eqnarray}
 \sqrt{Z_{\chi}} \chi_k^{\rm in}(t) & \sim & \frac{1}{\sqrt{2\omega_k(t)}}e^{-i\int^tdt\omega_k(t')}, \\
 \sqrt{Z_{\psi_{\chi}}} \psi_{\chi k}^{(+)s,{\rm in}}(t) & \sim &
  \frac{1}{\sqrt{2}}\cdot \sqrt{1+\frac{s|\mathbf{k}|}{\omega_k(t)}}e^{-i\int^tdt\omega_k(t')}, \\
 \sqrt{Z_{\psi_{\chi}}} \psi_{\chi k}^{(-)s,{\rm in}}(t) & \sim &
  \frac{\left<\phi(t)\right>}{\sqrt{2}\cdot \left|\left<\phi(t)\right>\right|}
  \sqrt{1-\frac{s|\mathbf{k}|}{\omega_k(t)}}e^{-i\int^tdt\omega_k(t')},
\end{eqnarray}
where $\omega_k \equiv \sqrt{\mathbf{k}^2+g^2\left|\left<\phi \right>\right|^2}$. WKB approximation is valid provided the condition $\left|\left<\phi \right>\right|\gg \sqrt{v/g}$ is fulfilled. Once one starts with a large enough vev $\left< \phi \right>$, one can safely choose for these solutions initial conditions in the form:
\begin{equation}
 \sqrt{Z_{\chi}} \chi_k^{\rm in}(0) = \frac{1}{\sqrt{2\omega_k(0)}},
  \qquad \dot{\chi}_k^{\rm in}(0) = -i\sqrt{\frac{\omega_k(0)}{2}},
\end{equation}
\begin{eqnarray}
 \sqrt{Z_{\psi_{\chi}}} \psi_{\chi k}^{(+)s,{\rm in}}(0) &=&
  \frac{1}{\sqrt{2}} \cdot \sqrt{1+\frac{s|\mathbf{k}|}{\omega_k(0)}}, \\
 \sqrt{Z_{\psi_{\chi}}} \psi_{\chi k}^{(-)s,{\rm in}}(0) &=&
  \frac{\phi(0)}{\sqrt{2}\cdot \left|\phi(0)\right|}\sqrt{1-\frac{s|\mathbf{k}|}{\omega_k(0)}}.
\end{eqnarray}
We can also assume WKB type solutions also for the out-state wave functions:
\begin{eqnarray}
 \sqrt{Z_{\chi}} \chi_k^{\rm out} &=& \frac{1}{\sqrt{2\omega_k}} e^{-i \int^t dt' \omega_k(t')}, \\
 \sqrt{Z_{\psi_{\chi}}} \psi_{\chi k}^{(+)s,{\rm out}} &=&
  \frac{1}{\sqrt{2}} \cdot \sqrt{1+\frac{sk}{\omega_k}} e^{-i \int^t dt' \omega_k(t')},\\
 \sqrt{Z_{\psi_{\chi}}} \psi_{\chi k}^{(-)s,{\rm out}} &=&
  \frac{\left< \phi \right>}{\sqrt{2} \left| \left< \phi \right> \right|}
  \cdot \sqrt{1-\frac{sk}{\omega_k}} e^{-i \int^t dt' \omega_k(t')}.
\end{eqnarray}
It is justified because we are still far away from the non-adiabatic region in the phase space.

Determining the vev of $\phi$ we consider the backreaction at the level of the asymptotic field expansion. Limiting ourselves to the 1-loop level in momentum integration we obtain the following equation fixing $\langle \phi \rangle$:
\begin{eqnarray}
 0 &=& \left< 0^{\rm in} \right| \left( \ \partial^2 \phi + g^2 \left| \chi \right|^2 \phi
  + \frac{1}{2}g \psi_{\chi}^{\dagger} \psi_{\chi}^{\dagger} \right) \left| 0^{\rm in} \right> \\
 & \sim & \left< \ddot{\phi} \right> + g \int \frac{d^3p}{(2\pi)^3} \left(
  Z_{\chi} \left| \chi_p^{\rm in} \right|^2 \cdot g \left< \phi \right>
  - \frac{1}{2} Z_{\psi_{\chi}} \sum_s \psi_{\chi p}^{(-)s, {\rm in}} \psi_{\chi p}^{(+)s,{\rm in}*} \right).
\end{eqnarray}
Putting together all pieces of information we can obtain approximate expressions for number densities of produced particles:
\begin{eqnarray}
 \left| \beta_{\chi k} \right|^2 & \sim &
  \frac{Z_{\chi} \left( \left| \dot{\chi}_k^{\rm in} \right|^2
   + \omega_k^2 \left| \chi_k^{\rm in} \right|^2 \right)}{2\omega_k} - \frac{1}{2}, \label{n_chi}\\
 \left| \beta_{\psi_{\chi}k}^s \right|^2 & \sim & \frac{1}{2} + \frac{sk}{2\omega_k} Z_{\psi_{\chi}}
   \left( \left| \psi_{\chi k}^{(-)s,{\rm in}} \right|^2 - \left| \psi_{\chi k}^{(+)s,{\rm in}} \right|^2 \right) \nonumber \\
  & & \qquad \qquad \quad - \frac{1}{\omega_k} Z_{\psi_{\chi}}{\rm Re}
    \left( g\left< \phi \right> \psi_{\chi k}^{(-)s,{\rm in}*} \psi_{\chi k}^{(+)s,{\rm in}} \right). \label{n_psi_chi}
\end{eqnarray}

\subsection{Numerical results} \label{sec:numerical_results}

\begin{figure}[t]
 \begin{center}
  \includegraphics[scale=0.7]{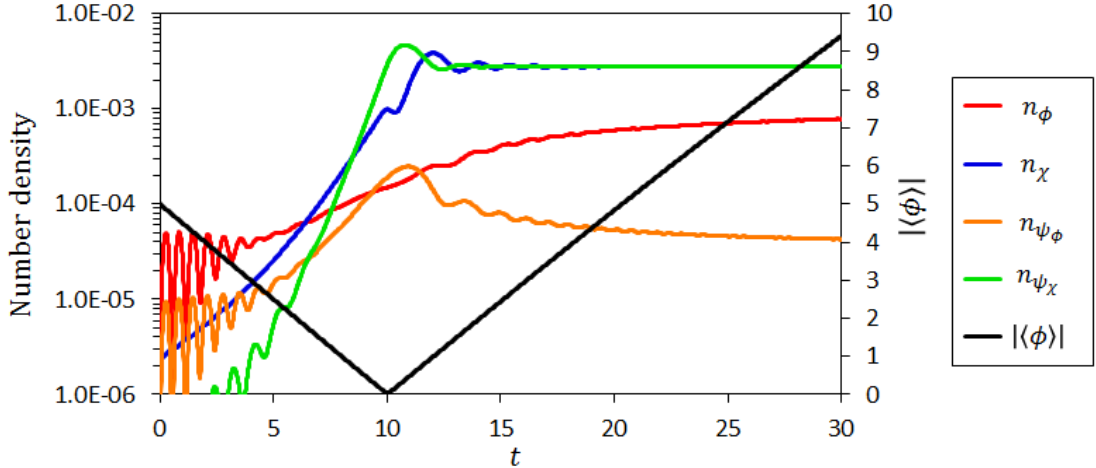}
  \caption{Picture shows our numerical time evolution of $|\langle \phi \rangle|$ (right vertical axis) and number densities (left vertical axis).
  We took $\phi(t=0)=5.0+0.05i$, $\dot \phi(t=0)=-0.5$, $g=1$.
  Values of number density at $t=30$ are: $n_{\phi}=7.82\times 10^{-4}$, $n_{\chi}=2.77\times 10^{-3}$,
  $n_{\psi_{\phi}}=4.26\times 10^{-5}$, $n_{\psi_{\chi}}=2.78\times 10^{-3}$.}
  \label{fig:number_density_g=1}
 \end{center}
\end{figure}

Equations describing number density of produced particles, (\ref{eq:phi_occu}), (\ref{eq:psi_phi_occu}), (\ref{n_chi}) and (\ref{n_psi_chi}), cannot be solved accurately analytically but are not so difficult to be solved numerically.
Numerical results for number densities are shown in the Figure \ref{fig:number_density_g=1}.
One can see that particle production occurrs not only for massive particles but also for the massless ones, as expected.
For the produced number of massive species, numerical results show a good agreement with analytic ones. For reasonable values of free parameters, $g=1, |v|=0.5, \mu=0.05$, one finds analytically: $n_{\chi} \sim n_{\psi_{\chi}} \sim 2.81 \times 10^{-3}$, what is in a good agreement with our numerical results. Comparing the massless particle number with massive one, for the above set of parameters,
the ratios after the period of production, at $t=30$, are $n_{\phi}/n_{\chi} \sim 28\%$ for bosons
and $n_{\psi_{\phi}}/n_{\psi_{\chi}} \sim 1.5 \%$ for fermions.

\begin{figure}[t]
 \begin{center}
  \includegraphics[scale=0.6]{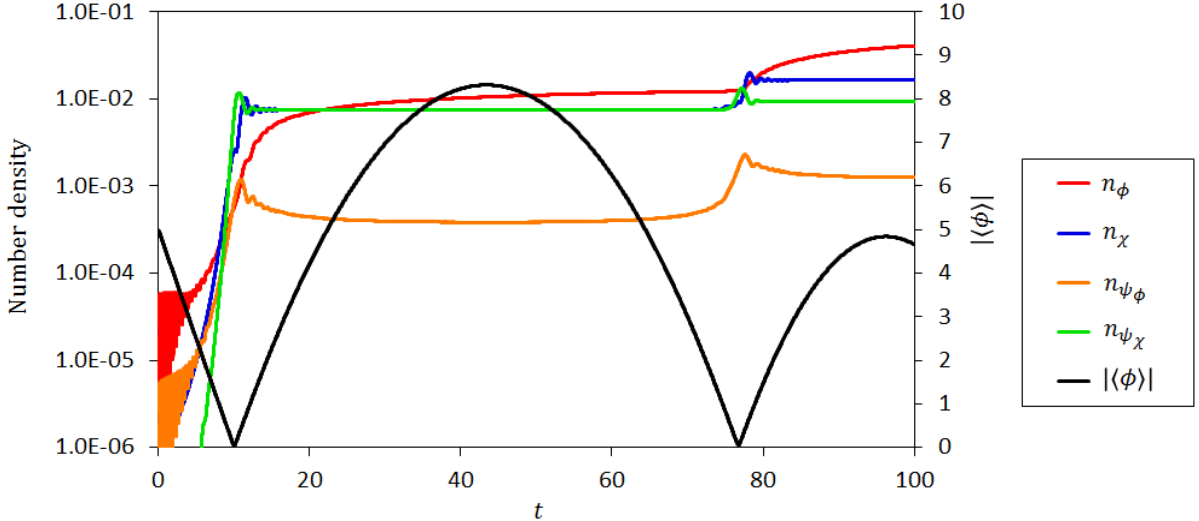}
  \caption{Time evolution of $|\langle \phi \rangle|$ and number densities in case of stronger coupling ($g=2$).
  Other parameters are the same as in the weaker case in Figure \ref{fig:number_density_g=1}.}
  \label{fig:number_density_g=2}
 \end{center}
\end{figure}

For comparison, in Figure \ref{fig:number_density_g=2} we also show our numerical results for stronger coupling. Weak and strong couplings have the usual meaning in this paper but crucial difference between them is that stronger coupling leads to the trapping effect mentioned earlier while weak coupling does not.
\noindent{In} the case of strong coupling, massless bosons $\phi$ are produced as abundantly as massive bosons $\chi$. One can also observe explicitly the second round of particle production which is caused by the trapping effect for $\left< \phi \right>$. 
One might wonder why the produced number density of $\phi$ becomes larger than that of massive particles. This may be an artefact caused by limited accuracy of numerical calculations (only up to terms of the order of $g^2$).
Nevertheless, it is safe to conclude  that massless particles can be produced as abundantly as massive particles provided the necessary couplings are sizeable.

\begin{figure}[t]
 \begin{center}
  \includegraphics[scale=0.5]{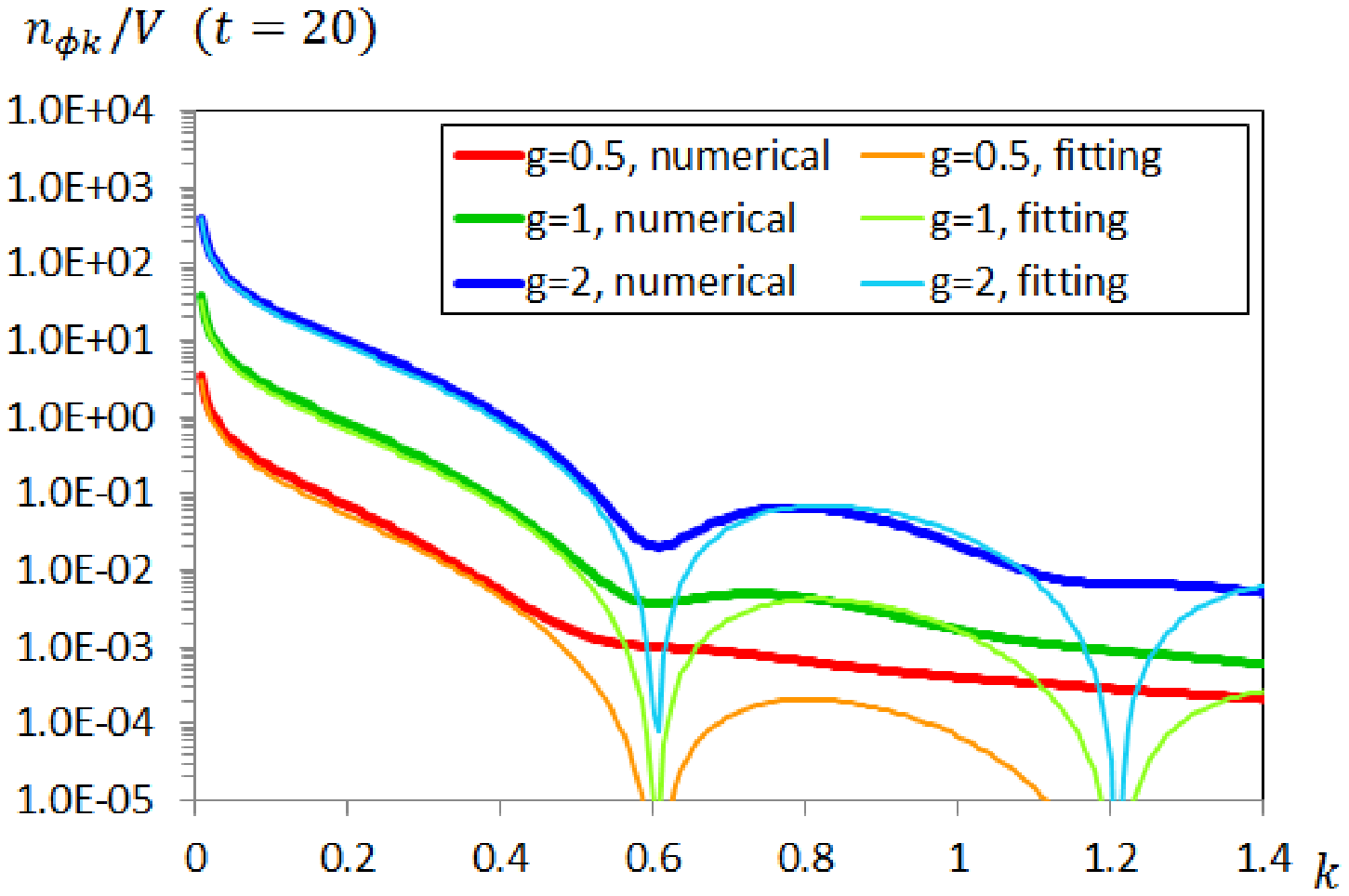}
  \includegraphics[scale=0.5]{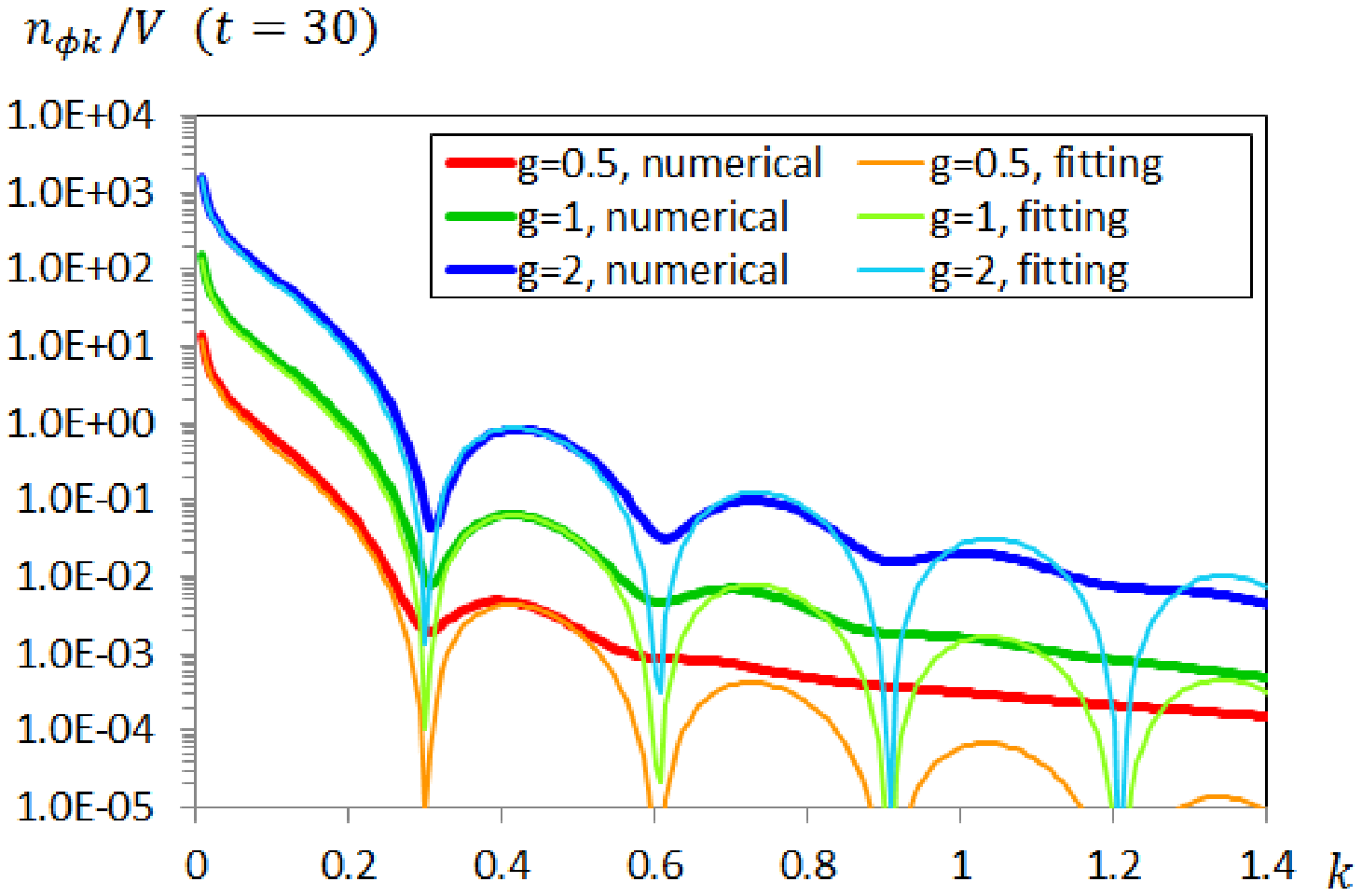}
  \includegraphics[scale=0.5]{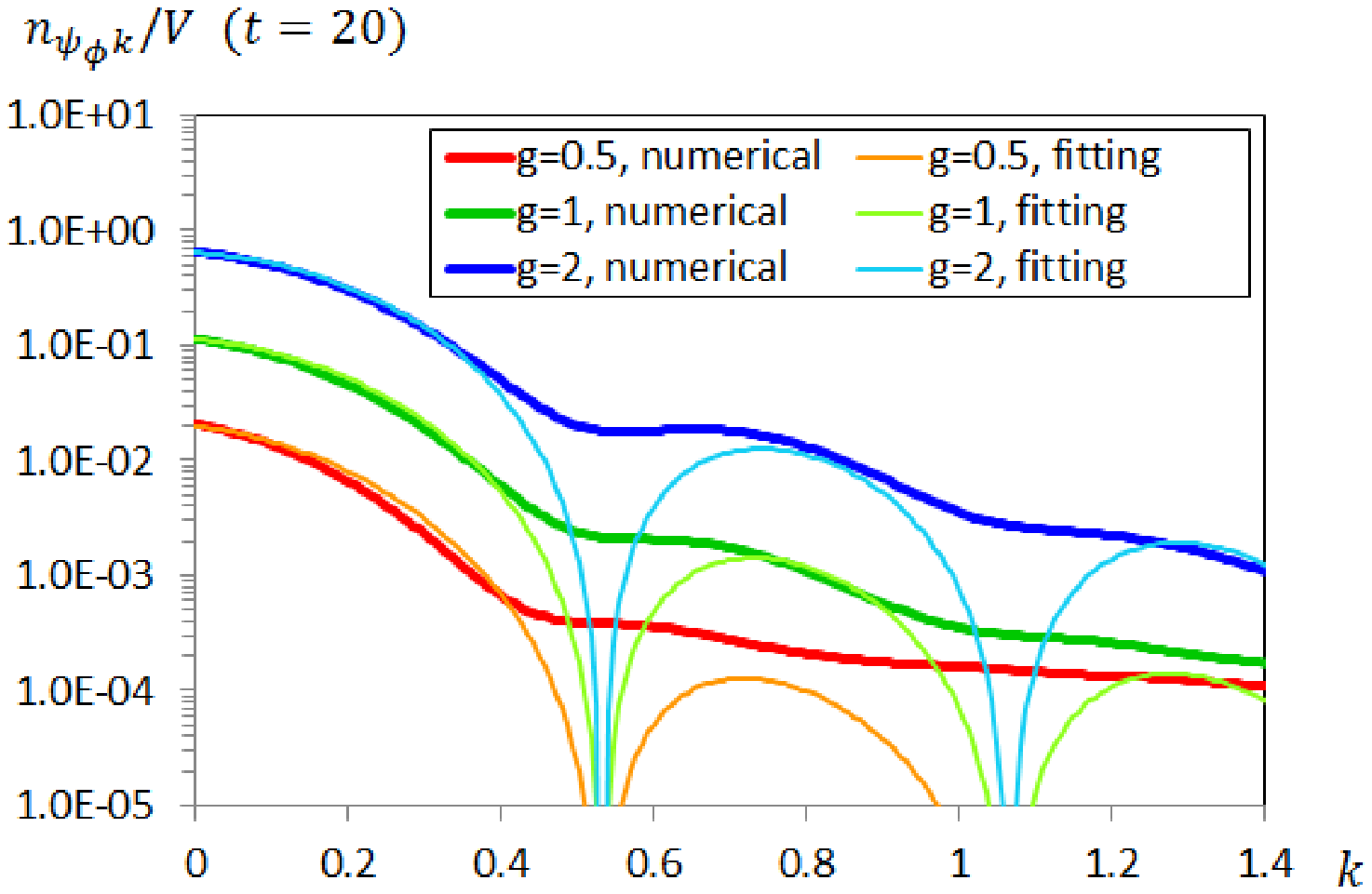}
  \includegraphics[scale=0.5]{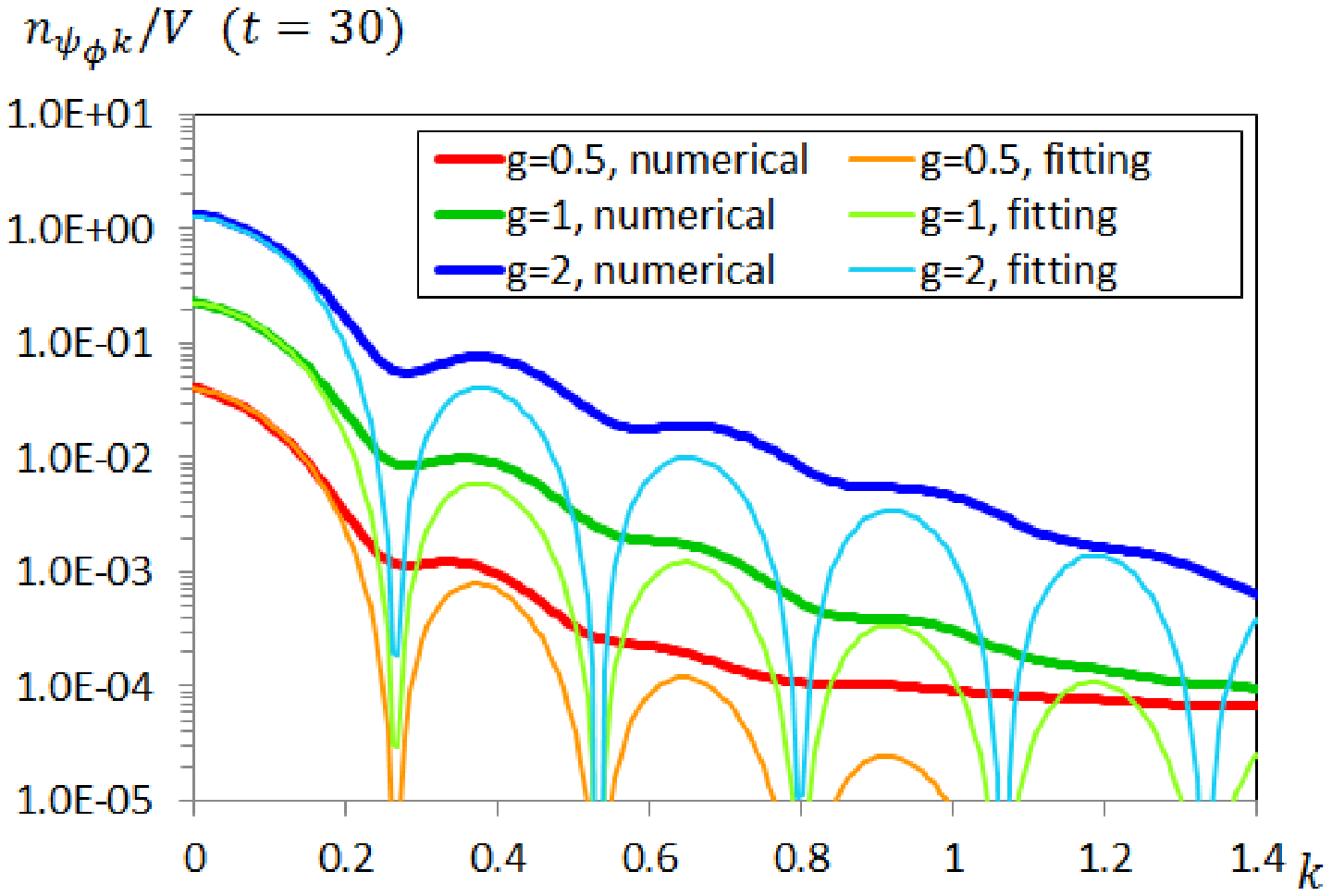}
  \caption{Numerical results of distribution functions for massless particles $n_{\phi k}, n_{\psi_{\phi}k}$ (thick lines)
  and their corresponding fitting functions for various couplings $g=0.5, 1$, and $2$ (thin lines).
  The initial values for numerical results are $\phi(t=0)=5.0+0.05i$ and $\dot \phi(t=0)=-0.5$ (in Figure \ref{fig:number_density_g=1}).
  For the fitting functions we set $|v|=0.5, t_*=10$ which are chosen to be consistent with initial conditions for numerical results.}
  \label{fig:fittings}
 \end{center}
\end{figure}

Finally, we show numerical results of distribution function for massless species and their fitting functions as in Figure \ref{fig:fittings}.
The fitting functions are chosen as follows:
\begin{eqnarray}
 n_{\phi k} / V & \sim & 0.16 \cdot \frac{g^2}{4\pi} \frac{1}{e^{\sqrt{\pi k^2/g|v|}}-1}
  \cdot g|v|(t-t_*)^2 \left[ \frac{\sin{0.52k(t-t_*)}}{0.52k(t-t_*)} \right]^2, \label{eq:n_phi_k_fitting} \\
 n_{\psi_{\phi} k} / V & \sim & 0.40  \cdot \frac{g^2}{4\pi} \frac{1}{e^{\sqrt{\pi k^2/g|v|}}+1}
  \cdot \sqrt{g|v|}(t-t_*) \left[ \frac{\sin{0.59k(t-t_*)}}{0.59k(t-t_*)} \right]^2, \label{eq:n_psi_phi_k_fitting}
\end{eqnarray}
where $t_*$ is the point of  time when a trajectory of $\phi$ is closest  to the  massless point ($\phi=0$).
Although these functions have been chosen in such a way that  their behaviour at small momentum $k$ satisfies (\ref{eq:n_phi_k_low_momemntum}) and (\ref{eq:n_psi_phi_k_low_momemntum}),
they follow very faithfully the numerical results over the whole available range of momenta due to the additional sine function.
The structure of the fitting functions is quite interesting.
Basically, they consist of the usual bosonic/fermionic distributions\footnote{It is interesting to notice that  the expression $\left[\exp{\left(\pi k^2/g|v|\right)} \pm 1 \right]^{-1}$ drops  at high momentum much faster than the numerical results.} multiplied by perturbative suppression $g^2/4\pi$. 

\section{Summary}
In this paper, we have described the effect of interactions in theories where the particle production is caused by a time-varying background field.
The necessary formalism has been introduced and analytic and numerical calculations have been performed in a simple but illustrative model.
As a general conclusion, we have shown that the amount of produced particles is related not only to the usual Bogoliubov coefficients $\beta_k$
but also to the magnitude and nature of the interaction terms.
In order to evaluate interaction effects and to compare their impact with a contribution coming from $\beta_k$,
we have considered a specific (supersymmetric) model which contains both massless fields $\phi, \psi_{\phi}$ ($\beta_k=0$)
and massive fields $\chi, \psi_{\chi}$ whose mass is varying in time due to a background field $\left< \phi \right>$ (hence these fields have $\beta_k \neq 0$).
As the result, we have found not only explicit results for massive particles but also quite interesting effects  for massless particles,
which are induced by interactions of massless fields with massive ones and, directly and indirectly, with the background field.
As illustrated by the fitting functions, the massless distribution is suppressed by a factor about $g^2/4\pi$ compared with the usual bosonic/fermionic distributions.
Since these results are derived by the perturbative method, they are not necessarily valid in the truly strong coupling region.
However, we can safely conclude that it is possible to create a sizeable amount of particles
with the help of interaction effects if the coupling is reasonably strong. \\

{\Large \bf Acknowledgements}\\

\noindent This work has been supported by the Polish NCN grant DEC-2012/04/A/ST2/00099.
\vspace*{0.3cm}
\appendix
\section{Helicity of fermions} \label{appen}

   In the fermionic deliberations we use the eigenvector of the helicity operator $e_{\mathbf{k}}^s$ ($s=\pm$) satisfying the relation
\begin{equation}
 k^i \bar{\sigma}^i e_{\mathbf{k}}^s = s |\mathbf{k}| \bar{\sigma}^0 e_{\mathbf{k}}^s. \nonumber
\end{equation}
Particular representation of this eigenvector is given by
\begin{equation}
 e_{\mathbf{k} \alpha}^s = \frac{1}{\sqrt{2}} \left(
 \begin{array}{c}
  \sqrt{1 - s k^3/|\mathbf{k}|} \\
  - s e^{i\theta_{\mathbf{k}}} \sqrt{1 + s k^3/|\mathbf{k}|}
 \end{array}
 \right)_{\alpha} , \nonumber
\end{equation}
which implies the following relations:
\begin{equation}
 e_{\mathbf{k} \alpha}^s e_{\mathbf{k} \dot{\alpha}}^{s \dagger}
  = \frac{1}{2} \left( \sigma^0 - \frac{s \sigma^i k^i}{|\mathbf{k}|} \right)_{\alpha \dot{\alpha}}, \:
 e_{\mathbf{k}}^{s \dagger \dot{\alpha}} e_{\mathbf{k}}^{s \alpha}
  = \frac{1}{2} \left( \bar{\sigma}^0 - \frac{s \bar{\sigma}^i k^i}{|\mathbf{k}|} \right)^{\dot{\alpha} \alpha}
  \: ({\rm no \: summation \: over \:} s). \nonumber
\end{equation}
In this paper we choose orthogonality conditions of the form
\begin{equation}
 e_{\mathbf{k}}^{s \dagger} \bar{\sigma}^0 e_{\mathbf{k}}^r =
  e_{\mathbf{k}}^r \sigma^0 e_{\mathbf{k}}^{s \dagger} = \delta^{sr}, \quad
 e_{\mathbf{-k}}^s e_{\mathbf{k}}^r = s e^{i\theta_{\mathbf{k}}} \delta^{sr}, \nonumber
\end{equation}
where $e^{i\theta_{\mathbf{k}}}$ is a phase factor defined as
\begin{equation}
 e^{i\theta_{\mathbf{k}}} \equiv \frac{k^1 + ik^2}{\sqrt{(k^1)^2 + (k^2)^2}}. \nonumber
\end{equation}

\end{document}